\documentclass[aps,prl,twocolumn,superscriptaddress,floatfix]{revtex4}

\usepackage{graphicx}
\begin{document}

%\preprint{}

\title{Reentrant Orbital Order and the True Ground State of
LaSr$_2$Mn$_2$O$_7$}

\author{Qing'An Li}
%\thanks{}
\affiliation{Chinese Academy of Sciences, Beijing, CHINA}
\affiliation{Materials Sciences Division, Argonne National Laboratory,
Argonne, IL 60439, USA}
\author{K.E. Gray}
\email[Electronic address: ]{KenGray@anl.gov}
\author{H. Zheng}
\author{H. Claus}
\author{S. Rosenkranz}
\author{S. Nyborg Ancona}
\author{R. Osborn}
\author{J.F. Mitchell}
\affiliation{Materials Sciences Division, Argonne National Laboratory,
Argonne, IL 60439, USA}
\author{Y. Chen}
\affiliation{NIST Center for Neutron Research, National Institute of
Standards and Technology, Gaithersburg, MD 20899, USA}
\affiliation{ Department of Materials Science and Engineering,
University of Maryland, College Park, MD 20742, USA}
\author{J.W. Lynn}
\affiliation{NIST Center for Neutron Research, National Institute of
Standards and Technology, Gaithersburg, MD 20899, USA}

\date{\today}

\begin{abstract}
Contrary to conventional wisdom, our purified
La$_{2-2x}$Sr$_{1+2x}$Mn$_2$O$_7$ crystals exhibit  {\it CE}-type
orbital and charge order as the low-temperature ground state for a hole
doping level $h$ = 0.5. For small deviations from $h$ = 0.5, the high
temperature {\it CE} phase is replaced at low temperatures by an
$A$-type antiferromagnet without coexistence. Larger deviations result
in a lack of  {\it CE} order at any temperature. Thus, small
inhomogeneities in cation or oxygen composition could explain why others
commonly see this reentrance with coexistence. 
\end{abstract}

\pacs{75.30.Vn; 75.30.Et; 75.60.-d}
%\keywords{}

\maketitle

The quest to better understand strongly correlated electrons is at the
heart of condensed matter inquiry. Colossal magnetoresistive manganites
exhibit a particularly vigorous competition among orbital, charge and
spin order \cite{jin94,moritomo96}. The phase diagrams, e.g.,
La$_{1-x}$Sr$_x$MnO$_3$ or the bilayer version
La$_{2-2x}$Sr$_{1+2x}$Mn$_2$O$_7$, display interesting features near
half doping ($x\sim0.5$) where, e.g., long-range orbital and
``checkerboard" charge ordering ( {\it CE} type) is predicted
\cite{goodenough55}. In bilayer manganites, it has been commonly
accepted that  {\it CE} order at $x$ = 0.5 is  reentrant
\cite{kubota99,ling00,argyriou00,chatterji00,kimura98,wakabayashi99,
wilkins03}: it forms below $\sim$210 K, but then is replaced by an {\it
A}-type antiferromagnet (AAFM) below $\sim$100 K. The lack of a
low-temperature  {\it CE} ordered ground state is surprising as it is
found at $x$ = 0.5 in many perovskite manganites \cite{tomioka}. In
bilayer manganites, coexistence of  {\it CE} order and AAFM between
$\sim$100 K and $\sim$200 K is also universally reported
\cite{kubota99,ling00,argyriou00,chatterji00,wakabayashi99}. 

All reports of reentrance in LaSr$_2$Mn$_2$O$_7$ are based on
superlattice peaks in neutron
\cite{kubota99,ling00,argyriou00,chatterji00}, x-ray
\cite{argyriou00,chatterji00,kimura98,wakabayashi99,wilkins03} and
electron \cite{li98,li01,luo05} diffraction and/or a peak in the
resistivity \cite{kubota99,argyriou00,kimura98,li98}. However,
consistent with our own experience, these reported properties for $x$ =
0.5 doping are variable. This implies an exquisite sensitivity to the
exact value of hole doping, $h = x - \delta$ in
La$_{2-2x}$Sr$_{1+2x}$Mn$_2$O$_{7-\delta}$. This dictates a vital burden
to obtain sample uniformity. Thus, in the present study we have purified
large single crystals, which often exhibit compositional gradients, by
cleaving them into very small crystals ($\sim$1 mg) and discarding those
that do not pass our test for uniformity (see below). Combining
conductivity, magnetization, and neutron and high-energy x-ray
diffraction data on such highly homogeneous crystals, we here show that
the  {\it CE} order predicted by Goodenough \cite{goodenough55} is the
low-temperature ground state, presumably at $h$ = 0.5, and that
coexistence of  {\it CE} and AAFM order is absent. We argue that
re-entrance only occurs for small deviations from 0.5 and propose a
revised phase diagram for bilayer manganites near $h$ = 0.5. 

Crystals were melt-grown in an optical image furnace \cite{mitchell97}.
The $c$ axis is perpendicular to the platelike crystals ($\sim$2
$\times$ 0.5 $\times$ 0.1 mm$^3$) and four gold pads are deposited along
the top and bottom surfaces for transport measurements. In order to
obtain a direct measure of the different order parameters, we performed
high-energy synchrotron X-ray scattering experiments at the 1-ID-C
station of the Advanced Photon Source at Argonne National Laboratory and
neutron scattering experiments at the BT-7 triple-axis instrument at the
NIST Center for Neutron Research. We utilize the (9/4,1/4,0) reflection
obtained using 80 keV X-rays and the neutron intensities of the
(1/4,1/4,3) and (1,1,3) reflections as measures of the  {\it CE} orbital
order, the  {\it CE} antiferromagnetic order (CEAFM), and the {\it
A}-type antiferromagnetic order, respectively \cite{kubota99}.

\begin{figure}[b]
\includegraphics[width=\columnwidth]{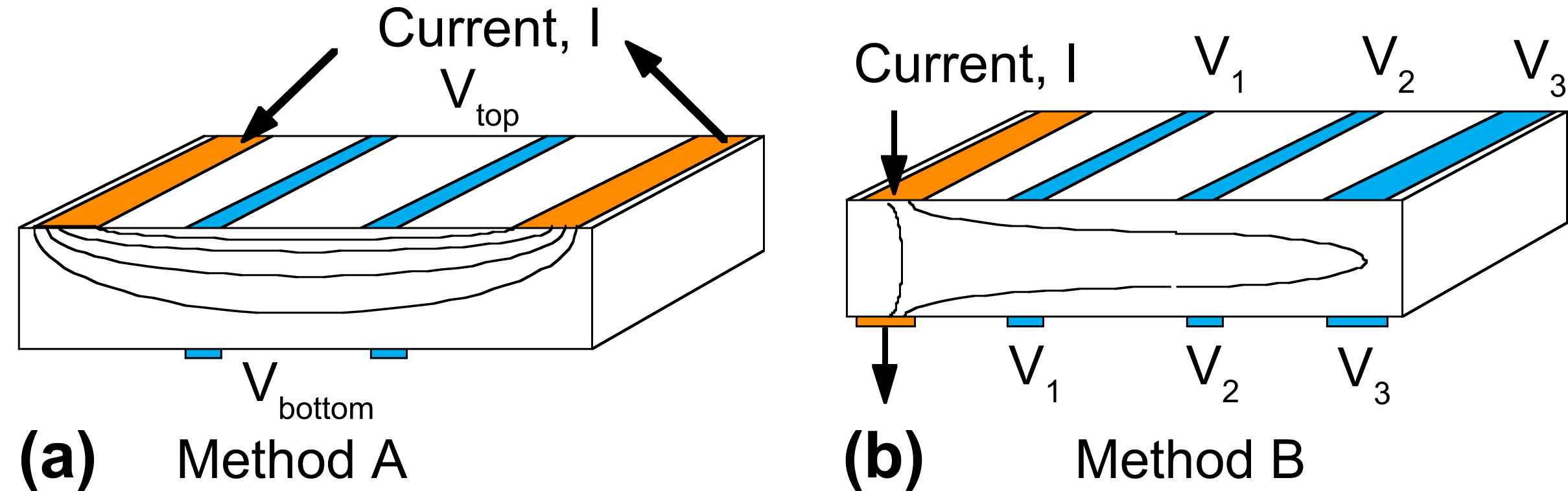}
\caption{\label{fig1}Schematic of six-terminal
configurations for method {\it A} (a) and method {\it B} (b). }
\end{figure}
Since four-terminal methods are unreliable for bulk crystals, we use six
terminals to determine each principal component of conductivity, i.e.,
along the $c$-axis, $\sigma_c$, and in the ab-plane, $\sigma_{ab}$
\cite{li99}. In transport method {\it A}, current is injected through
the outermost contacts on one surface [Fig.\ 1(a)]. Laplace's equation
is solved and inverted to get $\sigma_{ab}$ and $\sigma_c$ from voltages
measured across the innermost contacts of each surface. In method {\it
B}, Laplace's equation is solved for current injection through the top
and bottom contacts at one end of the crystal, while voltages are
measured between pairs of contacts on the top and bottom of the crystal
[Fig.\ 1(b)].

\begin{figure}[b]
\includegraphics[width=8.5cm]{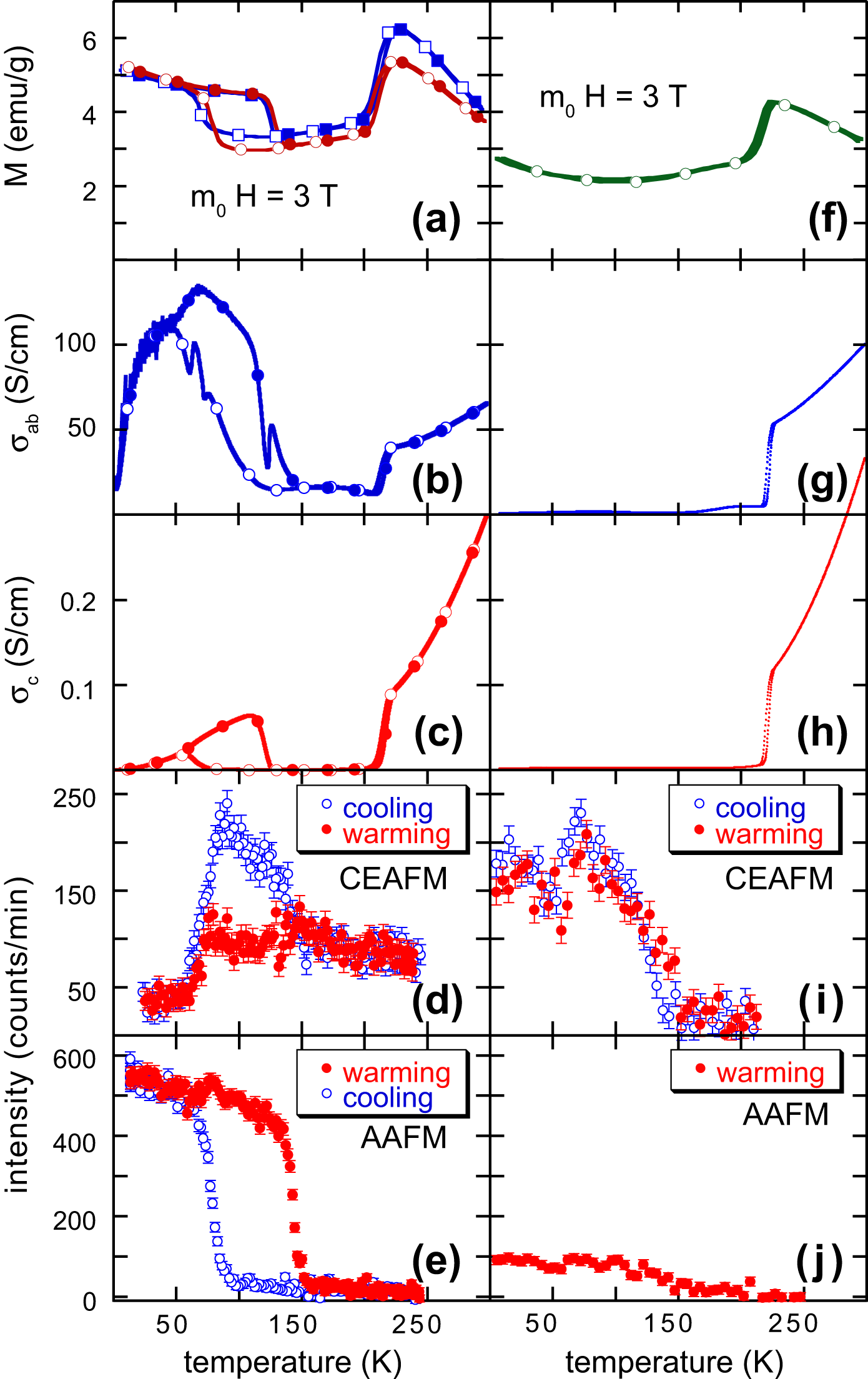}
\caption{\label{fig2} Temperature dependence, shown for
reentrant crystals (a)$-$(e) and nonreentrant crystals (f)$-$(j), of
magnetization (a),(f), conductivity in $ab$-plane (b),(g) and along c-axis
(c),(h) and neutron diffraction for CEAFM (d),(i) and {\it A}-type AFM (e),(j).
Open symbols refer to cooling and filled symbols to warming. }
\end{figure}

To test for homogeneity, we evaluate $\sigma_{ab}$ and $\sigma_c$ in
four configurations. Two use method {\it A} with current applied to the
outer contacts of either the top or bottom surfaces and two use method
{\it B} with current applied to the end contacts on either the right- or
left-hand sides. We require both $\sigma_{ab}(T)$ and $\sigma_c(T)$ to
be qualitatively the same for all four configurations and within a
factor of 2-3 in magnitude. Crystals used for Fig. 2 and 3 all pass our
criteria, as described in Ref.\ \cite{li06}. Homogeneous crystals with a
nominal $x$ = 0.5 fall into three batches: batch-{\it A} are  reentrant,
batch-{\it B} exhibits  {\it CE} order as the predominant
low-temperature ground state, and batch-{\it C} never exhibit  {\it CE}
order. Examples of these are indicated in the highly schematic phase
diagram of Fig.\ 4. The fact that we see three, and only three, {\it
unique} states using four different bulk probes is unmistakable evidence
of sufficient compositional uniformity. Any broad distribution of $h = x
- \delta$ would result in the coexistence of these states in some of the
12 crystals studied. The relevance of $\delta$ is seen in one reentrant
crystal that transformed into a non-reentrant crystal after annealing in
pure oxygen for 60 hrs at 600 $^\circ$C. Our scattering probes show that
our samples consist of up to two or three slightly misaligned
crystallites but are free from any impurity phases within our detection
limits.
\begin{figure}[b]
\includegraphics[width=8.5cm]{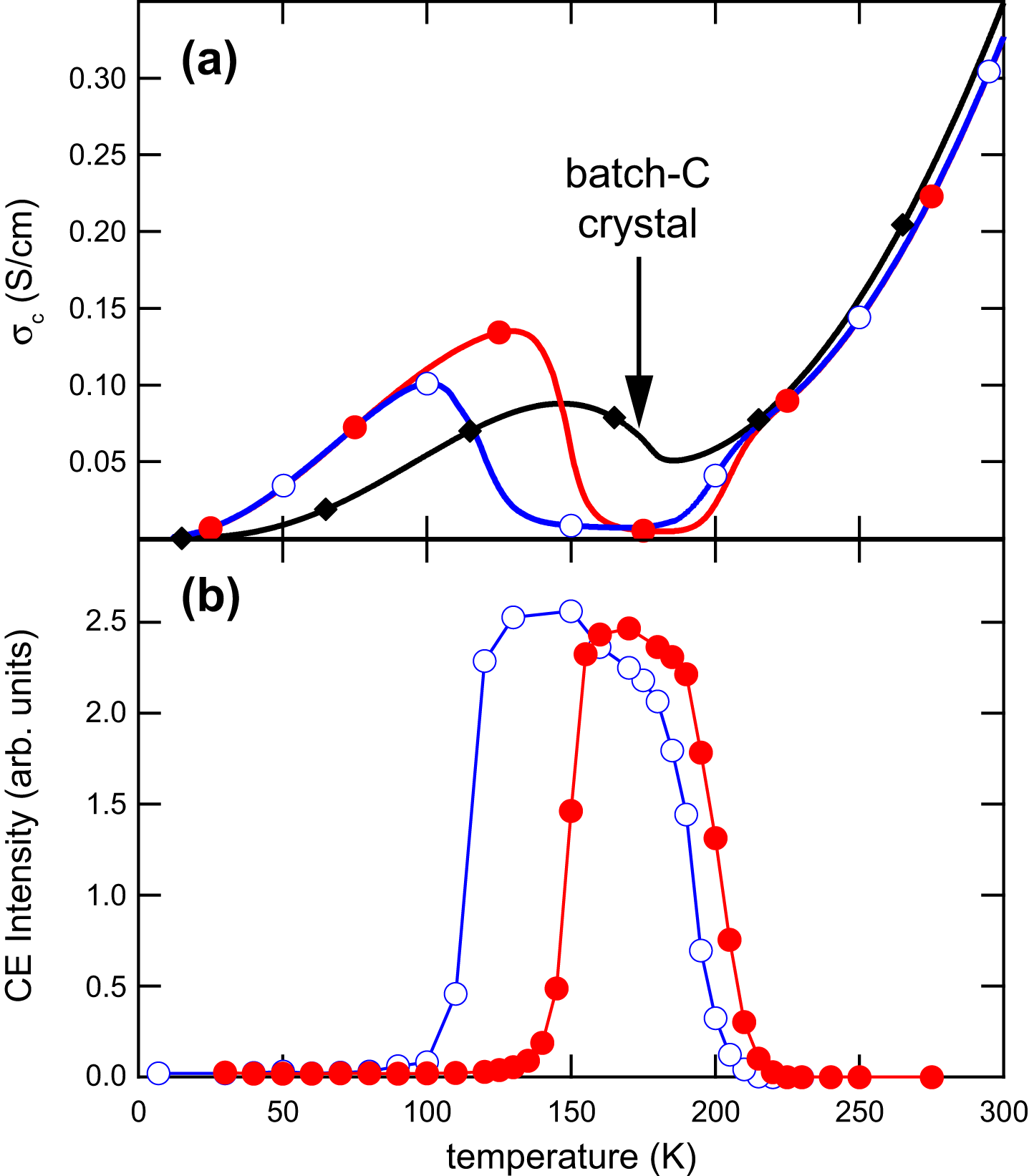}
\caption{\label{fig3} Temperature dependencies of (a)
$\sigma_c$ and (b) x-ray diffraction intensity of the superlattice
reflection for  {\it CE} order (9/4,1/4,0), measured on the same
crystal. Reentrance and hysteresis show a perfect correlation upon
cooling (open symbols) and warming (filled symbols). Also shown in (a)
are the conductivity data (black diamonds) for a batch-{\it C} crystal.
}
\end{figure}

Most of our crystals of nominal $x$ = 0.5 composition display a
transition from a paramagnetic insulator (PMI) above $\sim$200 K into a
{\it CE}-type orbital and charge ordered state. This is often
(batch-{\it A}) followed by a hysteretic transition into a state with
higher magnetization [Fig. 2(a)] and conductivity [Figs. 2(b) and 2(c)]
below $\sim$100 K that does not exhibit the  {\it CE}-type superlattice
reflections [Figs. 2(d) and 3]. Our neutron diffraction data [Fig. 2(e)]
confirm that the low-temperature state is the previously identified AAFM
\cite{kubota99,ling00,argyriou00,chatterji00}. However, in several
crystals (batch-{\it B}) no more than a few percent [Figs. 2(f)-2(j)]
transformed into the AAFM. A natural explanation is that  {\it CE} order
is the stable ground state only over a very narrow range of $h$ and
there are crystal-to-crystal variations of the average hole doping
$\langle h\rangle$ and a finite width of the distribution $\Delta h$.
Thus, when $\langle h\rangle$ is most favorable for  {\it CE} order,
presumably at 0.5 as predicted by Goodenough \cite{goodenough55}, a
larger fraction of the crystal would exhibit the  {\it CE} ground state
at low temperatures. Then for sufficiently small $\Delta h$, the  {\it
CE} ground state can be the majority phase. For crystals exhibiting
re-entrance we believe $\langle h\rangle$ differs somewhat from 0.5. We
also find crystals (batch-{\it C}) without a transition to the  {\it
CE}-state, implying that $\langle h\rangle$ is yet further from 0.5. The
conductivity data for batch-{\it C} crystals are very similar to that of
Ref.\ \cite{chen03} and are shown in Fig.\ 3(a) for a nominal doping $x$
= 0.48. Note that the sharp decrease in $\sigma_c$ at $\sim$ 200 K,
found in the reentrant and nonreentrant crystals, is entirely absent,
and the conductivity changes directly from the PMI to the AAFM behavior.

\begin{figure}[t]
\includegraphics[width=\columnwidth]{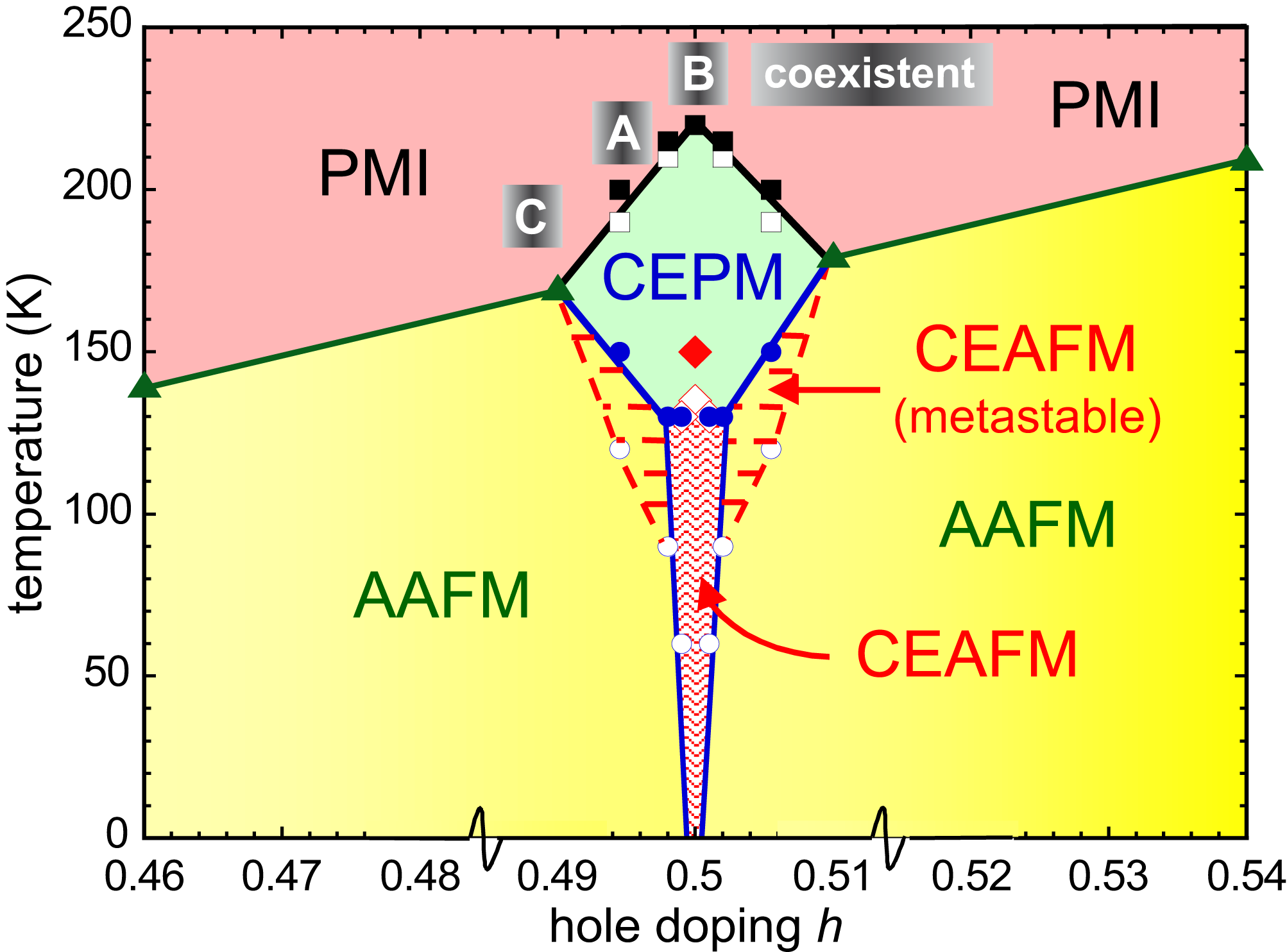}
\caption{\label{fig4} Schematic, qualitative phase diagram
near $h$ = 0.5. Symbols are measured transition temperatures (open:
cooling; filled: warming). Except for $h$ = 0.46 and 0.54, the $h$
values are arranged to connect smoothly with other data since we cannot
determine them with sufficient precision. Reentrant crystal data are
plotted symmetrically both above and below 0.5. The metastable CEAFM is
found on cooling but not on warming. The width of the boxes labeled
coexistent, {\it A}, {\it B}, and {\it C} represent suggested range of
$h$-values (i.e., $\langle h\rangle \pm \Delta h$) for coexistent
crystals and batches {\it A}, {\it B} and {\it C}. }
\end{figure}

Data on batch-{\it A} crystals display a striking temperature dependence
and hysteresis in $\sigma_{ab}$ and $\sigma_c$ and magnetization, as
shown in Figs.\ 2(a)-2(c) and 3(a). Similar hysteresis is also found in
the diffraction data of Figs.\ 2(d), 2(f) and 3(b) and in the data of
others \cite{kubota99,chatterji00}. Such hysteresis was also reported in
resonant x-ray studies \cite{wakabayashi99}, but we are unaware of
reports of such dramatic hysteresis in the conductivity or magnetization
of LaSr$_2$Mn$_2$O$_7$. The drops in our conductivity and magnetization
data exhibit a similar temperature range and hysteresis as all the
published diffraction data \cite{kubota99,chatterji00,wakabayashi99}. A
striking conclusion of our neutron diffraction data for this batch-{\it
A} crystal is the lack of coexistence of  {\it CE} and AAFM order, which
is universally found by others \cite{kubota99,argyriou00,chatterji00}.
Neutrons probe AAFM and CEAFM magnetic order [Figs.\ 2(d) and 2(e)], the
latter of which only occurs below $\sim$130 K (above 130 K the  {\it CE}
state is paramagnetic). In warming and cooling cycles, one, and only
one, state is ever found. This conclusion is possible because of the
small $\Delta h$ in our crystals. Near the lower-temperature transition
of batch-{\it A} crystals, we found evidence for sluggish kinetics that
was assisted by magnetic fields of 7 T. This may imply a small energy
difference between these states since the change in magnetization at the
transition is only $\sim$2 emu/g. Further, the broad thermal hysteresis
associated with reentrance implies there is only a slight difference in
the temperature dependencies of their free energies. 

Crystals from batch-{\it B} exhibit a similar high-temperature
transition, but only a trace of the lower temperature transition to AAFM
states. This is clearly seen in magnetization [Fig.\ 2(f)] and
conductivity [Figs.\ 2(g) and 2(h)]. Apparently the free energy for
{\it CE} order, that is the majority phase of batch-{\it B} crystals, is
sufficiently low so that  {\it CE} order remains stable down to low
temperatures. Since  {\it CE} order has a sharp minimum in its free
energy for $h$ exactly 0.5, batch-{\it B} crystals should have $\langle
h\rangle$ very close to that value. We confirm the low-temperature  {\it
CE} state through the CEAFM reflection [Fig.\ 2(i)] by neutron
diffraction in a slightly larger nonreentrant crystal. A minor part of
batch-{\it B} crystals may transform into the AAFM below $\sim$100 K
[Fig.\ 2(j)]. Others \cite{chatterji00,wilkins03} have identified a weak
{\it CE} superlattice peak at low temperatures which decays upon warming
to above $\sim$50 K, but then reappears as a strong peak at $\sim$120 K.
This has the appearance of a nonequilibrium ``quenched-in"  {\it CE}
state, that is then annealed out upon warming to $\sim$50 K. To dispel
this possibility and address thermodynamic stability, we monitor these
reflections during slow cooling ($\alt$ 1 K/min) that should minimize
quenched-in  {\it CE}-order due to sluggish kinetics. We find that the
CEAFM reflection is reversible upon slow heating [Fig.\ 2(i)], and
conclude that the  {\it CE} phase is the low temperature ground state in
a majority of each batch-{\it B} crystal.

For batch-{\it B} crystals, the succession of states with decreasing
temperature (shown schematically in Fig.\ 4) can be cast in terms of
entropy. Transforming from the PMI to a charge-ordered paramagnet (CEPM)
gains orbital and charge order while the onset of AF in the  {\it CE}
state (CEAFM) at $\sim$130 K additionally gains magnetic order.
Transitions between the AAFM and CEAFM states are more complex: the
internal energy for a CEAFM increases as $h$ deviates from 0.5 while the
broad hysteresis implies similar temperature dependenies of their free
energies. Therefore the phase boundary is almost vertical versus $h$. 
Batch-{\it A} crystals also transform from PMI to CEPM at $\sim$200 K
but  {\it CE} order appears to be metastable at lower temperatures
(dashed lines in Fig.\ 4) until it transforms at $\sim$100 K to the AAFM
ground state.
Within this metastable region, the  {\it CE} state develops CEAFM
magnetic order below $\sim$130 K due to a gain in magnetic entropy. Upon
warming, CEAFM order is absent and AAFM order persists until CEPM order
is thermodynamically stable (T $>$ 130 K). Thus it appears that the
magnetic entropy gain in the CEPM is necessary to overcome the barrier
between AAFM and  {\it CE} states upon warming. 

Batch-{\it C} crystals show no evidence in conductivity or high-energy
x-ray diffraction for  {\it CE} order although they were made with a
nominal composition of $x$ = 0.5. The  {\it CE} superlattice reflections
seen in batch-{\it A} and batch-{\it B} crystals were missing at all
temperatures down to 100 K. Curiously, all previously published
conductivity data \cite{kubota99,argyriou00,kimura98,tomioka} and in
particular \cite{chen03} known to us for nominal $x$ = 0.5 look more
like our batch-{\it C} crystals, although some show a small hysteresis
\cite{kubota99,argyriou00,kimura98}. A resistivity comparison
\cite{note} among these indicates a remarkable consistency in
temperature dependence and thus establishes our batch-{\it C} crystals
as a commonly seen variant of the nominal $x$ = 0.5 layered manganite.
The data of Refs.\ \cite{argyriou00,kimura98} do show a somewhat larger
resistivity peak at $\sim$180 K that could be the signature of some
{\it CE}-order in their crystals and each of these report the  {\it CE}
reflection. However, the resistivity peak associated with  {\it CE}
order in our batch-{\it A} crystals is at least 10 times larger than the
maximum value reported by others \cite{argyriou00,kimura98}. This may
imply  {\it CE} order occurs in a larger fraction of our purified
batch-{\it A} crystals. 

The electronic nature of AAFM states near $h$ = 0.5 is not so easily
determined. Conductivity for $x$ = 0.58 crystals indicate ab-plane
metals \cite{badica04} , whereas insulating behavior is seen for $x$ =
0.46 \cite{li03}. Conductivity data for the AAFM states of batch-{\it A}
and batch-{\it C} crystals fall between these two extremes. A possible
scenario is a continuous ab-plane-metal to insulator transition as $h$
decreases from $\sim$0.58 to $\sim$0.46. In this picture the  {\it CE}
order replaces the AAFM over a limited $h$-range centered at 0.5, but
the $h$-dependence of the AAFM states are otherwise unaffected. 

In summary, contrary to published data and accepted wisdom for
LaSr$_2$Mn$_2$O$_7$, we show that the zero-temperature ground state is
the  {\it CE} type predicted by Goodenough \cite{goodenough55}. We also
find no evidence of  {\it CE} and AAFM coexistence. That we do not know
the exact $h$-values is not critical to these two new observations, but
compositional purity was crucial to these discoveries. It was
accomplished by stringent testing of small ($\sim$1 mg) crystals and
verified by our observations with bulk probes of only three, {\it
unique} states in the 12 crystals tested. The lack of sufficient purity
could explain why others consistently see coexistence of  {\it CE} and
AAFM order.

\begin{acknowledgments}
The authors thank Peter Lee for assistance with synchrotron x-ray
diffraction at the Advanced Photon Source. This research was supported
by the U.S.\ Department of Energy, Basic Energy Sciences-Materials
Sciences, under contract No.\ DE-AC02-06CH11357.
\end{acknowledgments}


\end{document}